\documentclass[10pt]{amsart}
\usepackage{bbding}
\usepackage{stmaryrd}
\usepackage{amsfonts} 
\usepackage{graphicx,latexsym,bm,amsmath,amssymb,verbatim,multicol,lscape}
\newtheorem{thm}{Theorem} [section]

\newtheorem{lem}[thm]{Lemma}
\newtheorem{prop}[thm]{Proposition}
\newtheorem{prob}[thm]{Problem}
\theoremstyle{definition}

\theoremstyle{remark}

\newtheorem{con}[thm]{Conjecture}
\numberwithin{equation}{section}

\newcommand{\abs}[1]{\left\vert#1\right\vert}

\newcommand{\ra}{\rightarrow}

\newcommand{\F}{\mathbb{F}_2}
\newcommand{\cm}[1]{}
\newcommand{\m}[1]{\left\langle#1\right\rangle}
\begin{document}

\title{Nonlinearity of quartic rotation symmetric Boolean functions}
\author{Liping Yang}
\address{Mathematical College, Sichuan University, Chengdu 610064, P.R. China}
\email{lele20080823@yahoo.cn}
\author{Rongjun Wu}
\address{Mathematical College, Sichuan University, Chengdu 610064, P.R. China}
\email{eugen\_woo@163.com}
\author{Shaofang Hong*}
\address{Mathematical College, Sichuan University, Chengdu 610064, P.R. China
and Yangtze Center of Mathematics, Sichuan University, Chengdu 610064, P.R. China}
\email{sfhong@scu.edu.cn, s-f.hong@tom.com, hongsf02@yahoo.com}
\thanks{*Corresponding author and was supported partially by National
Science Foundation of China Grant \# 10971145 and by the Ph.D. Programs
Foundation of Ministry of Education of China Grant \#20100181110073}

\date{\today}%
\begin{abstract}
Nonlinearity of rotation
symmetric Boolean functions is an important topic on cryptography algorithm.
Let $e\ge 1$ be any given integer. In this paper, we investigate the
following question: Is the nonlinearity of the quartic rotation
symmetric Boolean function generated by the monomial $x_0x_ex_{2e}x_{3e}$
equal to its weight? We introduce some new simple sub-functions and develop
new technique to get several recursive formulas. Then we use these recursive
formulas to show that the nonlinearity of the quartic rotation
symmetric Boolean function generated by the monomial $x_0x_ex_{2e}x_{3e}$
is the same as its weight. So we answer the above question affirmatively.
Finally, we conjecture that if $l\ge 4$ is an integer, then the nonlinearity
of the rotation symmetric Boolean function generated by the monomial
$x_0x_ex_{2e}...x_{le}$ equals its weight.

\end{abstract}

\maketitle

\section{Introduction}

Rotation symmetric Boolean functions \cite{[CS2]} \cite{[SM]} are special kinds of
Boolean function whose evaluations on every cyclic inputs are the same.
We denote two vectors $(x_0, x_1, ...,\\ x_{n-1})$ and $(c_0, c_1, ..., c_{n-1})$
in $\F^n$ by $x$ and $c^n$ respectively, and define {\it dot product} of vectors by
$c^n \cdot x := \sum_{i=0}^{n-1}c_i x_i$. Given any vector $c^n$, we define
the linear function $l_{c^n}$ by $l_{c^n}(x):=c^n\cdot x$. By $\m{i}$
we denote the least nonnegative residue of $i \mod n$. We call the map
$f: \mathbb{F}_2^{n}\ra \mathbb{F}_2$ an {\it $n$ variables Boolean function},
and define its {\it weight} to be the number of
$x \in \F^n$ satisfying $f(x) = 1$, denoted by $wt(f)$.
For any two $n$ variables Boolean functions $f$ and $g$,
the distance $d(f, g)$ between $f$ and $g$ is defined by
$$
d(f, g):=wt(f-g)=\#\{a^n=(a_0, a_1, \ldots, a_{n-1}) \in
\mathbb{F}_2^n : f(a^n) \neq g(a^n)\}.
$$
We define the {\it Fourier transform} (sometimes called
{\it Walsh transform}) $\widehat{f^n}(c^n)$ of
the Boolean function $f^n$ at $c^n \in \mathbb{F}_2^n$ to be
$$
\widehat{f^n}(c^n)=\sum_{x\in
\mathbb{F}_2^n}(-1)^{f^n(x)+c^n\cdot x}.
$$
Evidently, we have
\begin{equation}
\widehat{f^n}(c^n)=2^n-2wt(f^n+l_{c^n}). \label{eq:ftc}
\end{equation}
In particular,
\begin{equation}
\widehat{f^n}(0)=2^n-2wt(f^n). \label{eq:ft0}
\end{equation}
The {\it nonlinearity} $N_{f^n}$ of $f^n$ is defined by
$$N_{f^n}:=\min \{d(f^n, l_{c^n}) \mid c^n \in \mathbb{F}_2^n\}.$$

In \cite{[KPH]}, Kim, Park and Hahn showed that the nonlinearity of
the quadratic rotation symmetric Boolean functions generated by
the monomial $x_0x_1$ may not be equal to its weight.
In \cite{[CS]}, Cusick and St\v{a}nic\v{a} conjectured that the nonlinearity
of the cubic rotation symmetric Boolean function in $n$ variables generated by
the monomial $x_0x_1x_2$ is the same as its weight. Ciungu \cite{[C]} and
Zhang, Guo, Feng and Li \cite{[ZGFL]} confirmed this conjecture. It is proved in
\cite{[ZGFL]} that the nonlinearity of the cubic rotation symmetric Boolean function
in $n$ variables generated by the monomial $x_0x_{e}x_{2e}$ with $e$
being a given positive integer equals its weight.
One naturally asks the following interesting question.

\noindent{\prob \label{prob} Let $e\geq 1$ be any given integer. Is the nonlinearity
of the quartic rotation symmetric Boolean function in $n$ variables
generated by the monomial $x_0x_{e}x_{2e}x_{3e}$ equal to its weight?}\\

In this paper, our main goal is to investigate these quartic rotation symmetric
Boolean functions. We introduce some new simple sub-functions and develop
new technique to show several recursive formulas. Then we use these
recursive formulas to show the following main result of this paper.

{\thm \label{thm:4rotsbf}
Let $e\geq 1$ be any given integer. Then the nonlinearity $N_{F_{4, e}^n}$ of
$$
F_{4, e}^n(x_0,...,x_{n-1}) := \sum_{i=0}^{n-1} x_{i} x_{\m{i+e}}
x_{\m{i+2e}}  x_{\m{i+3e}}
$$
is equal to its weight $wt(F_{4, e}^n)$.}\\

Evidently, Theorem \ref{thm:4rotsbf} answers Problem \ref{prob} affirmatively.
Letting $e=1$ in Theorem \ref{thm:4rotsbf} gives the main
result of \cite{[WZC]}. Note that the approach of \cite{[WZC]} is complicated and
completely similar to that of \cite{[ZGFL]} and that there is a common author of
\cite{[WZC]} and \cite{[ZGFL]} but \cite{[WZC]} did not cite \cite{[ZGFL]}. The method of
this paper is different from and simpler and much better than that of \cite{[WZC]}.
Actually, using the method of this paper, we can give a short proof of
the main result of \cite{[ZGFL]}. We have also produced more general results
by developing the method of the present paper.

This paper is organized as follows. In Section 2, we discuss Fourier
transform of Boolean functions. We will give some recursive formulas.
In Section 3, we give a sufficient condition such that the nonlinearity
of the rotation symmetric Boolean function is equal to its weight.
Using this result, we then prove Theorem \ref{thm:4rotsbf}. We propose
a conjecture at the end of this paper.

\section{Fourier transform of Boolean functions }

Throughout this paper, we let $F_4^n:=F_{4, 1}^n$.
First we introduce some notations. Let $i,j$ and $k$ be
nonnegative integers such that
$0 \le i \le j$ and $1 \le k \le j - i +1$. For brevity, we let $X(i,j,0):=0$,
$$
X(i,j,k):=\sum_{r=0}^{k-1} \prod_{s=i+r}^{j} x_s
$$
and
$$t_n:=\sum_{0 \leq i \leq n-4} x_i x_{i+1} x_{i+2} x_{i+3}.$$
Define
\begin{align*}
 &f_{0,0}^{n}(x):=t_{n},\\
 &f_{1,0}^{n}(x):=t_{n}+X(n-3,n-1,1),\\
 &f_{2,0}^{n}(x):=t_{n}+X(n-3,n-1,2),\\
 &f_{3,0}^{n}(x):=t_{n}+X(n-3,n-1,3).
\end{align*}
Evidently, if $x_{n-1}=0$, then $t_{n-1}=t_n$ and $X(n-3,n-1,i)=0$. This implies that
\begin{equation} \label{eq:0}
t_n+X(n-3,n-1,i)=t_{n-1}
\end{equation}
for $0\le i\le 3$ if $x_{n-1}=0$.

Also if $x_{n-1}=1$, then
\begin{equation} \label{eq:1}
t_n+X(n-3,n-1,i)=
\left\{
  \begin{array}{ll}
    t_{n-1}+X(n-4,n-2,i+1), & \hbox{if $i=0,1,2$,} \\
    t_{n-1}+X(n-4,n-2,3)+1, & \hbox{if $i=3$.}
  \end{array}
\right.
\end{equation}

For $0\le i\le 3$, let
\begin{align*}
&f_{i,1}^{n}(x):=f_{i,0}^{n}(x)+x_0 x_1 x_2,\\
&f_{i,2}^{n}(x):=f_{i,0}^{n}(x)+x_0 x_1 x_2+x_0 x_1,\\
&f_{i,3}^{n}(x):=f_{i,0}^{n}(x)+x_0 x_1 x_2+x_0 x_1+x_0.
\end{align*}

It is obvious that $f_{i,j}^{n}(x_0, ..., x_{n-1}) = f_{j,i}^{n}(x_{n-1},...,x_0)$
and that $\widehat{f_{i,j}^{n}}(0) = \widehat{f_{j,i}^{n}}(0)$.

Now we have
\begin{equation} \label{eq:F4n}
\begin{split}
\widehat{F_4^n}(c^n)&
=(1+(-1)^{c_{n-2}})\widehat{f_{0,0}^{n-3}}(c^{n-3})
    +(-1)^{c_{n-1}}\widehat{f_{0,1}^{n-3}}(c^{n-3})\\
+&(-1)^{c_{n-2}+c_{n-1}}\widehat{f_{0,2}^{n-3}}(c^{n-3})
+(-1)^{c_{n-3}}        \widehat{f_{1,0}^{n-3}}(c^{n-3})\\
+&(-1)^{c_{n-3}+c_{n-1}}\widehat{f_{1,1}^{n-3}}(c^{n-3})
+(-1)^{c_{n-3}+c_{n-2}}\widehat{f_{2,0}^{n-3}}(c^{n-3})\\
+&(-1)^{c_{n-3}+c_{n-2}+c_{n-1}}\widehat{f_{3,3}^{n-3}}(c^{n-3}).
\end{split}
 \end{equation}
Particularly, we have
\begin{equation} \label{eq:F4n0}
\begin{split}
\widehat{F_4^n}(0)
=2\widehat{f_{0,0}^{n-3}}(0)
    + \widehat{f_{0,1}^{n-3}}(0)
+ \widehat{f_{0,2}^{n-3}}(0)
+  \widehat{f_{1,0}^{n-3}}(0)
+ \widehat{f_{1,1}^{n-3}}(0)
+ \widehat{f_{2,0}^{n-3}}(0)
+ \widehat{f_{3,3}^{n-3}}(0).
\end{split}
 \end{equation}

We give recursive relations about $\widehat{f_{i,j}^{n}}(c^n)$ for $ 0 \le i,j \le 3$.

{\lem \label{lm:1} Let $0\leq j\leq 3$ be an integer, for every $c^n=(c_0, ..., c_{n-1}) \in \F^n$ with $c_{n-1}=0$, we have
\begin{align*}
\widehat{f_{0,j}^n}(c^n) &= 2\widehat{f_{0,j}^{n-2}}(c^{n-2}) + 2(-1)^{c_{n-2}}\widehat{f_{0,j}^{n-3}}(c^{n-3}) +
                            2(-1)^{c_{n-2}+c_{n-3}}\widehat{f_{0,j}^{n-4}}(c^{n-4}),\\
\widehat{f_{1,j}^n}(c^n) &= 2\widehat{f_{0,j}^{n-2}}(c^{n-2}) + 2(-1)^{c_{n-2}}\widehat{f_{0,j}^{n-3}}(c^{n-3})+
                            2(-1)^{c_{n-2}+c_{n-3}+c_{n-4}}\widehat{f_{3,j}^{n-4}}(c^{n-4}),\\
\widehat{f_{2,j}^n}(c^n) &= 2\widehat{f_{0,j}^{n-2}}(c^{n-2})+
                            2(-1)^{c_{n-2}+c_{n-3}}\widehat{f_{0,j}^{n-4}}(c^{n-4}),\\
\widehat{f_{3,j}^n}(c^n) &= 2(-1)^{c_{n-2}}\widehat{f_{0,j}^{n-3}}(c^{n-3})+
                            2(-1)^{c_{n-2}+c_{n-3}+c_{n-4}}\widehat{f_{3,j}^{n-4}}(c^{n-4}),
\end{align*}
where $c^{i}$ is the vector consisting of the first $i$ bits of $c^n \in \F^n$ for $i=n-2,n-3 \text{ and } n-4$.
}
\begin{proof}
We give the proof of the recurrence equation for $\widehat{f_{0,0}^n}(c^n)$,
the same argument leads to the proof of others.
By Equation (\ref{eq:0}) and (\ref{eq:1}), we have

\begin{align*}
\widehat{f_{0,0}^n}(c^n) &= (\sum_{x:x_{n-1}=0}+\sum_{x:x_{n-1}=1}) (-1)^{t_n+c^n\cdot x}\\
                         &= \sum_{x^{n-1}} (-1)^{t_{n-1}+c^{n-1}\cdot x^{n-1}}
                            + \sum_{x^{n-1}} (-1)^{t_{n-1}+X(n-4,n-2,1)+c^{n-1}\cdot  x^{n-1}}\\
                         &= 2 \sum_{x^{n-2}} (-1)^{t_{n-2}+c^{n-2}\cdot  x^{n-2}}
                            + \sum_{x^{n-2}} (-1)^{t_{n-2}+X(n-5,n-3,1)+c_{n-2}+c^{n-2}\cdot  x^{n-2}}\\
                       + & \sum_{x^{n-2}} (-1)^{t_{n-2}+X(n-5,n-3,2)+c_{n-2}+c^{n-2}\cdot  x^{n-2}}\\
                         &= 2 \widehat{f_{0,0}^{n-2}}(c^{n-2}) + (-1)^{c_{n-2}}(2 \sum_{x^{n-3}} (-1)^{t_{n-3}+c^{n-3}\cdot  x^{n-3}}\\
                       + & \sum_{x^{n-3}} (-1)^{t_{n-3}+X(n-6,n-4,2)+c_{n-3}+c^{n-3}\cdot  x^{n-3}}
                            + \sum_{x^{n-3}} (-1)^{t_{n-3}+X(n-6,n-4,3)+c_{n-3}+c^{n-3}\cdot  x^{n-3}})\\
                         &=  2 \widehat{f_{0,0}^{n-2}}(c^{n-2}) + 2 (-1)^{c_{n-2}}  \widehat{f_{0,0}^{n-3}}(c^{n-3})
                       + 2 (-1)^{c_{n-2}+c_{n-3}}  \widehat{f_{0,0}^{n-4}}(c^{n-4}).
\end{align*}
The proof of Lemma \ref{lm:1} is complete.
\comment{The relations of $f_{1,0}$,$f_(2,0)$ and $f_(3,0)$ have been double checked.}
\end{proof}
{\lem \label{lm:2} Let $0 \le j \le 3$ be an integer. For every $c^n=(c_0, ..., c_{n-1}) \in \F^n$ with $c_{n-1}=1$, we have
\begin{align}
\nonumber \widehat{f_{0,j}^{n}}(c^n) &= (-1)^{c_{n-2}} \widehat{f_{1,j}^{n-2}}(c^{n-2}) + (-1)^{1+c_{n-2}} \widehat{f_{2,j}^{n-2}}(c^{n-2}),\\
\nonumber \widehat{f_{1,j}^{n}}(c^n) &= (-1)^{c_{n-2}} \widehat{f_{1,j}^{n-2}}(c^{n-2}) + (-1)^{1+c_{n-2}} \widehat{f_{3,j}^{n-2}}(c^{n-2}),\\
\widehat{f_{2,j}^{n}}(c^n) &= (-1)^{c_{n-2}} \widehat{f_{1,j}^{n-2}}(c^{n-2}) + (-1)^{c_{n-2}}   \widehat{f_{3,j}^{n-2}}(c^{n-2}),\\
\nonumber \widehat{f_{3,j}^{n}}(c^n) &=              2 \widehat{f_{0,j}^{n-2}}(c^{n-2}) +
                            2(-1)^{c_{n-2}+c_{n-3}}\widehat{f_{0,j}^{n-4}}(c^{n-4}),
\end{align}
where $c^{i}$ is the first $i$ bits of $c^n \in \F^n$ for $i=n-2,n-3 \text{ and } n-4$.
}
\begin{proof}
We only prove the recursive formula for $\widehat{f_{2,0}^{n}}(c^n)$
because the recursive formulas for others can be proved in the similar way.
Since $c_{n-1}=1$, it follows from (2.1) and (2.2) that
\begin{align*}
\widehat{f_{2,0}^{n}}(c^n)
&=\sum_{x}(-1)^{t_n+X(n-3,n-1,2)+c^n\cdot x}\\
&=\sum_{x}(-1)^{t_n+X(n-3,n-1,3)+c^{n-1}\cdot x^{n-1}}\\
&=(\sum_{x:x_{n-1}=0}+\sum_{x:x_{n-1}=1}) (-1)^{t_n+X(n-3,n-1,3)+c^{n-1}\cdot x^{n-1}}\\
&=\sum_{x^{n-1}}(-1)^{t_{n-1}+c^{n-1}\cdot  x^{n-1}}
    + \sum_{x^{n-1}}(-1)^{t_{n-1}+X(n-4,n-2,3)+1+c^{n-1}\cdot x^{n-1}}\\
&=\sum_{x^{n-2}}(-1)^{t_{n-2}+c^{n-2}\cdot  x^{n-2}}
    + \sum_{x^{n-2}}(-1)^{t_{n-2}+X(n-5,n-3,1)+c_{n-2}+c^{n-2}\cdot x^{n-2}}\\
&+\sum_{x^{n-2}}(-1)^{t_{n-2}+1+c^{n-2}\cdot  x^{n-2}}
    + \sum_{x^{n-2}}(-1)^{t_{n-2}+X(n-5,n-3,3)+c_{n-2}+c^{n-2}\cdot x^{n-2}}\\
&=(-1)^{c_{n-2}} \widehat{f_{1,0}^{n-2}}(c^{n-2}) + (-1)^{c_{n-2}}   \widehat{f_{3,0}^{n-2}}(c^{n-2}).
\end{align*}
Thus Lemma \ref{lm:2} is proved.
\end{proof}

{\lem \label{cor:1} Let $n \ge 8$ and $ 0 \le i,j \le 3$ be integers. Then we have
\begin{equation}\label{eq:2}
\widehat{f_{i,j}^n}(0) = 2 \widehat{f_{i,j}^{n-2}}(0)
                       + 2 \widehat{f_{i,j}^{n-3}}(0)
                       + 2 \widehat{f_{i,j}^{n-4}}(0).
\end{equation}
}
\begin{proof}
Using Maple 14, we can compute and obtain the results listed in Table \ref{tbf3}.
From Table \ref{tbf3}, we know that (2.6) is true when $8\le n\le 11$.
In the following we let $n\ge 12$.

By Lemma \ref{lm:1}, we have
\begin{equation*} 
\widehat{f_{0,j}^n}(0) = 2 \widehat{f_{0,j}^{n-2}}(0)
                       + 2 \widehat{f_{0,j}^{n-3}}(0)
                       + 2 \widehat{f_{0,j}^{n-4}}(0).
\end{equation*}
Thus (\ref{eq:2}) is true if $i=0$ and $0\le j\le 3$.  On the other hand,
it follows from (2.5) that
\begin{align*}
&\widehat{f_{2,j}^{n-2}}(0)+\widehat{f_{2,j}^{n-3}}(0)+\widehat{f_{2,j}^{n-4}}(0)\\
=& 2 (\widehat{f_{0,j}^{n-4}}(0)+\widehat{f_{0,j}^{n-5}}(0) +  \widehat{f_{0,j}^{n-6}}(0) )
 + 2 (\widehat{f_{3,j}^{n-6}}(0) + \widehat{f_{3,j}^{n-7}}(0)) + \widehat{f_{3,j}^{n-8}}(0))\\
=& \widehat{f_{0,j}^{n-2}}(0) + \widehat{f_{0,j}^{n-4}}(0)=\frac{1}{2}\widehat{f_{2,j}^n}(0).
\end{align*}
So (2.6) is proved when $i=2$ and $0\le j\le 3$.
In what follows we treat the remaining cases $i=1$ and 3.

First let $i=3$. We assume that (2.6) is true for all the $\le n-1$ case when $i=3$.
By Lemma \ref{lm:1}, the result for the $i=0$ case and the induction hypothesis, we derive that
\begin{align*}
&\widehat{f_{3,j}^{n-2}}(0)+\widehat{f_{3,j}^{n-3}}(0)+\widehat{f_{3,j}^{n-4}}(0)\\
=& 2 (\widehat{f_{0,j}^{n-5}}(0) +  \widehat{f_{0,j}^{n-6}}(0) + \widehat{f_{0,j}^{n-7}}(0))
 + 2 (\widehat{f_{3,j}^{n-6}}(0) + \widehat{f_{3,j}^{n-7}}(0) + \widehat{f_{3,j}^{n-8}}(0))\\
=& \widehat{f_{0,j}^{n-3}}(0) + \widehat{f_{3,j}^{n-4}}(0)=\frac{1}{2}\widehat{f_{3,j}^n}(0).
\end{align*}
Thus (2.6) is proved if $i=3$.

Finally, we let $i=1$. Clearly Lemma 2.1 tells us that
\begin{align*}
\widehat{f_{1,j}^{n-2}}(0) &= 2 (\widehat{f_{0,j}^{n-4}}(0) + \widehat{f_{0,j}^{n-5}}(0) + \widehat{f_{3,j}^{n-6}}(0)),\\
\widehat{f_{1,j}^{n-3}}(0) &= 2 (\widehat{f_{0,j}^{n-5}}(0) + \widehat{f_{0,j}^{n-6}}(0) + \widehat{f_{3,j}^{n-7}}(0)),\\
\widehat{f_{1,j}^{n-4}}(0) &= 2 (\widehat{f_{0,j}^{n-6}}(0) + \widehat{f_{0,j}^{n-7}}(0) + \widehat{f_{3,j}^{n-8}}(0)).
\end{align*}
Hence by the result for the cases $i=0$ and 3 and Lemma 2.1 we obtain that
\begin{align*}
&\widehat{f_{1,j}^{n-2}}(0)+\widehat{f_{1,j}^{n-3}}(0)+\widehat{f_{1,j}^{n-4}}(0)\\
=& 2 (\widehat{f_{0,j}^{n-4}}(0)+\widehat{f_{0,j}^{n-5}}(0) + \widehat{f_{0,j}^{n-6}}(0))
 + 2 (\widehat{f_{0,j}^{n-5}}(0)+\widehat{f_{0,j}^{n-6}}(0) + \widehat{f_{0,j}^{n-7}}(0))\\
+& 2 (\widehat{f_{3,j}^{n-6}}(0)+\widehat{f_{3,j}^{n-7}}(0) + \widehat{f_{3,j}^{n-8}}(0))\\
=&\widehat{f_{0,j}^{n-2}}(0)+\widehat{f_{0,j}^{n-3}}(0) +
\widehat{f_{3,j}^{n-4}}(0)=\frac{1}{2}\widehat{f_{1,j}^n}(0).
\end{align*}
Therefore (2.6) is true if $i=3$. This completes the proof of Lemma \ref{cor:1}.

\begin{table}
\centering \caption{The evaluations of $\widehat{f_{i,j}^{n}}(0)$, $0 \le i\leq j \le 3$
and $\widehat{F_4^n}(0)$ with respect to $ 4 \le n <12$.}\label{tbf3}
\begin{tabular}{c|rrrrrrrr}
 \hline
$n$                          & 4 & 5 &  6 &  7 &  8 &  9  & 10 &11\\
\hline\hline
$\widehat{f_{0,0}^{n}}(0)$   & 14&28&52&100&188&360&680&1296\\

$\widehat{f_{0,1}^{n}}(0)$   & 14&24&48&88&172&320&616&1160\\

$\widehat{f_{0,2}^{n}}(0)$   & 10&20&36&72&132&256&480&920\\

$\widehat{f_{0,3}^{n}}(0)$   & 6&8&20&32&68&120&240&440\\
\hline
$\widehat{f_{1,1}^{n}}(0)$   & 10&24&40&84&148&296&544&1056\\

$\widehat{f_{1,2}^{n}}(0)$   & 10&16&36&60&124&224&440&816\\

$\widehat{f_{1,3}^{n}}(0)$   & 2&12&12&36&52&120&200&416\\
\hline
$\widehat{f_{2,2}^{n}}(0)$   & 6&16&24&52&92&184&336&656\\

$\widehat{f_{2,3}^{n}}(0)$   & 6&4&16&20&52&80&176&304\\
\hline
$\widehat{f_{3,3}^{n}}(0)$   & -2&8&0&20&12&56&64&176\\
\hline
$\widehat{F_4^n}(0)$         & 16&20&52&84&176&312&624&1144\\
\hline
\end{tabular}
\end{table}

\end{proof}


{\thm \label{thm:2.4} Let $n \ge 8$ be an integer. Then we have
\begin{equation*}
\widehat{F_4^n}(0) = 2 \widehat{F_4^{n-2}}(0) + 2 \widehat{F_4^{n-3}}(0)
                         + 2 \widehat{F_4^{n-4}}(0).
\end{equation*}
}
\begin{proof}
By Table \ref{tbf3}, we know that Theorem 2.4 holds if $8\le n\le 11$. On the other hand,
Lemma \ref{cor:1} applied to (\ref{eq:F4n0}) gives us the desired result if $n\ge 12$.
So Theorem 2.4 is proved.
\end{proof}

\begin{table}
\centering \caption{The evaluations of $\widehat{f_{i,j}^{5}}(c^5)$, $0 \le i,j \le 3$,
as $c^5=(c_0,...c_4) \in \mathbb{F}_2^5$ with $c_1 \ne 0$ is denoted by $\sum_{0 \le i \le 4}c_i 2^i$. }\label{tbfijc}
$$
\begin{tabular}{c|rrrr|rrrr|rrrr|rrrr}
\hline
$c$&$\negthickspace\widehat{f_{0,0}^{5}}$&$\negmedspace\widehat{f_{0,1}^{5}}$
&$\negmedspace\widehat{f_{0,2}^{5}}$&$\negmedspace\widehat{f_{0,3}^{5}}$
&$\negthickspace\widehat{f_{1,0}^{5}}$&$\negmedspace\widehat{f_{1,1}^{5}}$
&$\negmedspace\widehat{f_{1,2}^{5}}$&$\negmedspace\widehat{f_{1,3}^{5}}$
&$\negthickspace\widehat{f_{2,0}^{5}}$&$\negmedspace\widehat{f_{2,1}^{5}}$
&$\negmedspace\widehat{f_{2,1}^{5}}$&$\negmedspace\widehat{f_{2,3}^{5}}
$&$\negthickspace\widehat{f_{3,0}^{5}}$&$\negmedspace\widehat{f_{3,1}^{5}}$
&$\negmedspace\widehat{f_{3,2}^{5}}$&$\negmedspace\widehat{f_{3,3}^{5}}$\\
\hline
\hline
2&4&8&12&-8&0&0&8&-12&4&8&8&-4&0&-4&4&-8\\
3&0&-4&-8&12&-4&-4&-12&8&0&-4&-4&8&-4&0&-8&4\\
\hline
6&-4&-8&4&-8&0&0&8&-4&-4&-8&0&-4&0&4&4&0\\
7&0&4&-8&4&4&4&-4&8&0&4&-4&0&4&0&0&4\\
\hline
10&-4&0&-4&0&0&8&0&4&-4&0&0&-4&0&4&-4&8\\
11&0&-4&0&-4&4&-4&4&0&0&-4&-4&0&4&0&8&-4\\
\hline
14&4&0&4&0&0&-8&0&-4&4&0&8&-4&0&-4&-4&0\\
15&0&4&0&4&-4&4&-4&0&0&4&-4&8&-4&0&0&-4\\
\hline
18&0&-4&0&-4&4&4&4&0&0&-4&4&-8&4&8&8&-4\\
19&-4&0&-4&0&0&0&0&4&-4&0&-8&4&0&-4&-4&8\\
\hline
22&0&4&0&4&-4&-4&-4&0&0&4&4&0&-4&-8&0&-4\\
23&4&0&4&0&0&0&0&-4&4&0&0&4&0&4&-4&0\\
\hline
26&0&4&0&4&-4&-4&-4&0&0&4&-4&8&-4&0&0&-4\\
27&4&0&4&0&0&0&0&-4&4&0&8&-4&0&-4&-4&0\\
\hline
30&0&-4&0&-4&4&4&4&0&0&-4&-4&0&4&0&8&-4\\
31&-4&0&-4&0&0&0&0&4&-4&0&0&-4&0&4&-4&8\\
\hline
\end{tabular}
$$
\end{table}

\section{Proof of theorem \ref{thm:4rotsbf}}

In the present section, we prove Theorem 1.2. We begin with the following result.
{\prop \label{prop1.3} If
$$
\widehat{f^n}(0)=\max \{ |\widehat{f^n}(c^n)| \mid c^n \in \mathbb{F}_2^n\},
$$
then we have $N_{f^n}=wt(f^n)$.}
\begin{proof}

Since
$
\widehat{f^n}(0)=\max \{ |\widehat{f^n}(c^n)| \mid c^n \in \mathbb{F}_2^n\},
$
we have
$$\min \{2^n - \widehat{f^n}(c^n) \mid  c^n \in \mathbb{F}_2^n\}=2^n-\widehat{f^n}(0).$$

By (\ref{eq:ftc}) and (\ref{eq:ft0}), we infer that
$$
2 wt(f^n+l_{c^n})=2^n-\widehat{f^n}(c^n)
$$
and
$$
2wt(f^n)=2^n-\widehat{f^n}(0).
$$
It follows that
\begin{equation} \label{eq:min}
\min\{wt(f^n+l_{c^n}) \mid c^n \in \mathbb{F}_2^n\}= wt(f^n).
\end{equation}
But
$$d(f^n, l_{c^n})=wt(f^n-l_{c^n})=wt(f^n+l_{c^n}).$$
Then we have
$$N_{f^n}=\min \{wt(f^n+l_{c^n}) \mid c^n \in \mathbb{F}_2^n \}.$$
Therefore, by (\ref{eq:min}) we deduce that $N_{f^n}=wt(f^n).$
This ends the proof of Proposition \ref{prop1.3}.
\end{proof}

{\lem \label{lm:key} Let $n \ge 4$ be an integer and $c^n=(c_0, ..., c_{n-1}) \in \F^n$.
If $c_1\ne 0$, then for all $0 \le i,j \le 3$, we have}
$$
|\widehat{f_{i,j}^n}(c^n)| < \frac{1}{8}\widehat{F_4^{n+3}}(0).
$$

\begin{proof}
We use induction on $n$. For all $ 4 \le n \le 8$ and $c^n$ with $c_1\neq 0$, using Maple 14,
we can verify that  $|\widehat{f_{i,j}^n}(c^n)| < \frac{1}{8}\widehat{F_4^{n+3}}(0)$
for all $0 \le i, j \le 3$. (For example, the case for $n=5$ is given in Table \ref{tbfijc},
from which we can read that $|\widehat{f_{i,j}^5}(c^5)| < 22= \frac{1}{8}\widehat{F_4^{8}}(0)$
for all $c^5$ with $c_1\ne 0$.)

Let $n\ge 9$. Assume that Lemma 3.2 is true for all $\le n-1$ case.
In what follows we prove Lemma 3.2 for the $n$ case.
Since $c_1 \neq 0$, we have $c^{i}$ is nonzero for all $n-4\le i\le n$.
We consider the following cases.

{\bf Case 1}. $c_{s-1}=0$. Then by Lemma \ref{lm:1}, the induction hypothesis and Theorem \ref{thm:2.4},
we deduce that for $0\le j\le 3$,

\begin{align*}
\abs{\widehat{f_{0,j}^n}(c^n)} &= \abs{2\widehat{f_{0,j}^{n-2}}(c^{n-2}) + 2(-1)^{c_{n-2}}\widehat{f_{0,j}^{n-3}}(c^{n-3}) +
                            2(-1)^{c_{n-2}+c_{n-3}}\widehat{f_{0,j}^{n-4}}(c^{n-4})}\\
                               &\leq 2\abs{\widehat{f_{0,j}^{n-2}}(c^{n-2})} + 2 \abs{\widehat{f_{0,j}^{n-3}}(c^{n-3})} +
                            2 \abs{\widehat{f_{0,j}^{n-4}}(c^{n-4})}\\
                               &<\frac{1}{8}(2 \widehat{F_4^{n+1}}(0) + 2 \widehat{F_4^{n}}(0)
                         + 2 \widehat{F_4^{n-1}}(0))\\
                               &=\frac{1}{8}\widehat{F_4^{n+3}}(0).
\end{align*}
Similarly, for $1 \le i \le 3$ and $0 \le j \le 3$, we can prove that if $c_{s-1}=0$, then
$|\widehat{f_{i,j}^n}(c^n)| < \frac{1}{8}\widehat{F_4^{n+3}}(0)$.

{\bf Case 2}. $c_{s-1}=1$. By Lemma \ref{lm:2}, the induction hypothesis and Theorem \ref{thm:2.4},
we derive that for $0 \le j \le 3$
\begin{align*}
\abs{\widehat{f_{3,j}^n}(c^n)} &= \abs{2 \widehat{f_{0,j}^{n-2}}(c^{n-2}) +
                                2(-1)^{c_{n-2}+c_{n-3}}\widehat{f_{0,j}^{n-4}}(c^{n-4})}\\
                               &\le 2 \abs{\widehat{f_{0,j}^{n-2}}(c^{n-2})} +
                                2\abs{\widehat{f_{0,j}^{n-4}}(c^{n-4})}\\
                               & < \frac{1}{8}(2 \widehat{F_4^{n+1}}(0) + 2 \widehat{F_4^{n-1}}(0))\\
                               & < \frac{1}{8}\widehat{F_4^{n+3}}(0).
\end{align*}
Similarly, for $0\le i \le 2$ and $0 \le j \le 3$, we can prove that if $c_{s-1}=1$, then
$|\widehat{f_{i,j}^n}(c^n)| < \frac{1}{8}\widehat{F_4^{n+3}}(0)$. The proof of Lemma 3.2 is complete.
\end{proof}


We can now prove Theorem \ref{thm:4rotsbf}.\\

\noindent {\it Proof of Theorem \ref{thm:4rotsbf}.}
First we show that Theorem 1.2 is true for the case $e=1$.
To do so, by Proposition \ref{prop1.3}
it is sufficient to show that
\begin{equation*}
\widehat{F_4^n}(0)=\max\{ |\widehat{F_4^n}(c^n)|
\mid  c^n \in \mathbb{F}_2^n\} \label{eq:main4}.
\end{equation*}
First from (\ref{eq:F4n}) we can derive that
\begin{equation*}
\begin{split}
|\widehat{F_4^n}(c^n)|
&\le 2|\widehat{f_{0,0}^{n-3}}(c^{n-3})|
    +|\widehat{f_{0,1}^{n-3}}(c^{n-3})|
+|\widehat{f_{0,2}^{n-3}}(c^{n-3})|
    +| \widehat{f_{1,0}^{n-3}}(c^{n-3})|\\
&+|\widehat{f_{1,1}^{n-3}}(c^{n-3})|
    +|\widehat{f_{2,0}^{n-3}}(c^{n-3})|
+|\widehat{f_{3,3}^{n-3}}(c^{n-3})|.
\end{split}
 \end{equation*}
By the definition of $F_4^n$, we can easily see that for any $0\le j\le n-1$,
$$\widehat{F_4^n}(c_0 ,..., c_{n-1})=\widehat{F_4^n}(c_j ,..., c_{\m{n-1+j}}).$$
For any $c^n\in \mathbb{F}_2^n$ with $c^n \neq 0$, without loss of any generality,
we can assume that $c_1 \neq 0$.
It then follows immediately from Lemma \ref{lm:key} that
 \begin{equation*}
\begin{split}
|\widehat{F_4^n}(c^n)|
&<\frac{1}{8}(2\widehat{F_4^n}(0)+\widehat{F_4^n}(0)+\widehat{F_4^n}(0)
+\widehat{F_4^n}(0)+\widehat{F_4^n}(0)+\widehat{F_4^n}(0)+\widehat{F_4^n}(0))\\
&=\widehat{F_4^n}(0).
\end{split}
 \end{equation*}
Thus Theorem \ref{thm:4rotsbf} is proved for the case $e=1$.

In what follows, we let $e>1$.
We define the permutation $\rho$ acting on index set $\{0,1, ..., n-1\}$ by
$$
\rho(i):= \m{i + e} , i \in \{0,1, ..., n-1\}.
$$
Then $\rho$ can be decomposed as follows.
\begin{align*}
\rho=& \left(
     \begin{array}{cccc}
       0 & 1     & \cdots & n-1 \\
       e & \m{e+1} & \cdots & e-1
     \end{array}
     \right)\\
     =& \left(
     \begin{array}{cccc}
       0 & e   & \cdots & \m{(t-1)e+k} \\
       e & \m{2e} & \cdots & 0
     \end{array}
     \right)
     \cdots \\
     &\left(
     \begin{array}{cccc}
       s-1       & \m{s+e-1}   & \cdots & \m{s+(t-1)e-1} \\
       \m{s+e-1} & \m{s+2e-1}  & \cdots & s-1
     \end{array}
     \right)\\
     =&\pi_0 \pi_1 \cdots \pi_{s-1},
\end{align*}
where
$$
\pi_k = \left(
     \begin{array}{cccc}
       k       & \m{e+k}   & \cdots & \m{(t-1)e+k} \\
       \m{e+k} & \m{2e+k}  & \cdots & k
     \end{array}
     \right) (0 \le k \le s-1)
$$
is the cycle permutation acting on $\{k, \m{e+k}, ..., \m{(t-1)e+k}\}$, $s=\gcd(n,e)$ and $ t=n/s$.
Then we derive that
\begin{align*}
F_{4, e}^n(x_0,...,x_{n-1})&=\sum_{i=1}^{n-1} x_i x_{\m{i + e}} x_{\m{i + 2 e}} x_{\m{i + 3 e}}\\
                      &=\sum_{0\leq k\leq{s-1}}\sum_{0\leq j\leq {t-1}}
                      x_{\m{k+je}}x_{\m{k+je+e}}x_{\m{k+je+2e}}x_{\m{k+je+3e}}\\
                      :&=\sum_{0\leq k\leq{s-1}} g_k^t(x_k,x_{\m{e+k}},...,x_{\m{(t-1)e+k}}).
\end{align*}

Through indeterminacies substitution
$$
x_{k+js} \rightarrow y_j^{(k)}, 0\leq k\leq {s-1},0\leq j\leq {t-1},
$$
we obtain
$$
g_k^t(x_k,x_{\m{e+k}},...,x_{\m{(t-1)e+k}})
=\sum_{0\leq j\leq {t-1}}y_j^{(k)}y_{\m{j+1}}^{(k)}y_{\m{j+2}}^{(k)}y_{\m{j+3}}^{(k)}.
$$
Let $c_k^t=(c_k,c_{\m{e+k}},...,c_{\m{(t-1)e+k}})$ and $x_k^t=(x_k,x_{\m{e+k}},...,x_{\m{(t-1)e+k}})$.
For any $c_k^t\neq0$, it follows immediately from Theorem \ref{thm:4rotsbf} for the case $e=1$ that
$$
\widehat{g_k^t}(c_k^t) < \widehat{g_k^t}(0).
$$
Then for all $c^n=(c_0, ..., c_{n-1}) \ne 0$, by definition of Fourier transform, we obtain that
\begin{align*}
&|\widehat{F_{4, e}^n}(c^n)|\\&=|\sum_{0\leq i \leq {n-1}}
(-1)^{x_i x_{\m{i+e}} x_{\m{i+2e}} x_{\m{i+3e}} + c^n\cdot x}|\\
&=|\prod_{0\leq i \leq {s-1}}\sum_{0\leq i \leq {t-1}}
(-1)^{x_{\m{k+je}}x_{\m{k+je+e}}x_{\m{k+je+2e}}x_{\m{i+je+3e}}+c^n\cdot x}|\\
&=|\prod_{0\leq i \leq {s-1}}\widehat{g_k^t}(c_k^t)|\\
&<\prod_{0\leq i \leq {s-1}}\widehat{g_k^t}(0)=\widehat{F_{4, e}^n}(0).
\end{align*} 
Therefore Theorem \ref{thm:4rotsbf} is true for the case $e>1$.

This concludes the proof of Theorem \ref{thm:4rotsbf}.
 \hfill$\Box$

\section{A conjecture}
Let $e\ge 1$ and $l\ge 2$ be integers and define the Boolean function
$F_{l, e}^n(x_0,...,x_{n-1})$ by
$$
F_{l, e}^n(x_0,...,x_{n-1}):= \sum_{i=0}^{n-1} x_{i} x_{\m{i+e}}
x_{\m{i+2e}}...x_{\m{i+(l-1)e}}.
$$
Kim, Park and Hahn \cite{[KPH]} showed that the nonlinearity $N_{F_{2, e}^n}$ of
$F_{2, e}^n$ may not be equal to its weight $wt(F_{2, e}^n)$. It is proved in \cite{[ZGFL]}
that $N_{F_{3, e}^n}=wt(F_{3, e}^n)$. By Theorem 1.2, we know that $N_{F_{4, e}^n}
=wt(F_{4, e}^n)$. For the general integer $l\ge 5$, we believe that such result should
be true. That is, we propose the following conjecture as the conclusion of this paper.
 
{\con \label{con}
Let $e\geq 1$ and $l\ge 5$ be any given integer. Then the nonlinearity $N_{F_{l, e}^n}$
of $F_{l, e}^n(x_0,...,x_{n-1})$ is equal to its weight $wt(F_{l, e}^n)$.}\\

By using and developing the method of this paper, one can confirm 
the above Conjecture 4.1 if $l=5$ and 6. However, we meet obstruction
when we try to prove Conjecture 4.1 for the general case that $l\ge 7$.

\end{document}